\documentclass[letterpaper, 10 pt, conference]{ieeeconf}  

\IEEEoverridecommandlockouts                              

\let\labelindent\relax
\usepackage{cite}
\usepackage{amsmath,amssymb,amsfonts}
\usepackage{algorithmic}
\usepackage{graphicx}
\usepackage{textcomp}
\usepackage{xcolor}
\usepackage{tabularray}
\usepackage{longtable}
\usepackage{tabularx}
\usepackage{fancyhdr}
\usepackage[utf8]{inputenc} 
\usepackage{gensymb}
\usepackage{enumitem}
\usepackage{lineno}
\usepackage{float} 
\usepackage{multirow}
\usepackage{longtable}
\usepackage{booktabs}
\usepackage{array}
\usepackage{amsmath, array,graphicx}
\usepackage{caption,subcaption}
\usepackage{nameref}
\usepackage{amsmath,amssymb}
\usepackage{gensymb}
\usepackage{tabularray}
\usepackage{adjustbox}
\usepackage{multirow}
\usepackage[pagebackref=true,breaklinks=true,letterpaper=true,colorlinks,bookmarks=true]{hyperref}

\newcommand{\AK}[1]{{\color{black}{#1}}}

\title{\LARGE \bf Early and Accurate Detection of Tomato Leaf Diseases Using TomFormer }

\author{Asim Khan$^{1}$, Umair Nawaz$^{2}$, Lochan Kshetrimayum$^{1}$, Lakmal Seneviratne$^{1}$, and Irfan Hussain$^{1}$
\thanks{*This research is supported by ASPIRE, the technology program management pillar of Abu Dhabi’s Advanced Technology Research Council (ATRC), under the ASPIRE project “Aspire Research Institute for Food Security in the Drylands '' within Theme 1.4.}
\thanks{$^{1}$A. Khan, L. Kshetrimayum, L. Seneviratne and I. Hussain are with the Department of Mechanical Engineering, Khalifa University, Abu Dhabi, United Arab Emirates.
        {\tt\small [@ku.ac.ae]}}%
\thanks{$^{2}$U. Nawaz Researcher is with the Department of Electrical Engineering, Namal University, Mianwali, Punjab, Pakistan        
        {\tt\small [1601005@namal.edu.pk]}}%
}

\begin{document}

\maketitle


\begin{abstract}

Tomato leaf diseases pose a significant challenge for tomato farmers, resulting in substantial reductions in crop productivity. The timely and precise identification of tomato leaf diseases is crucial for successfully implementing disease management strategies. This paper introduces a transformer-based model called TomFormer for the purpose of tomato leaf disease detection. The paper's primary contributions include the following: Firstly, we present a novel approach for detecting tomato leaf diseases by employing a fusion model that combines a visual transformer and a convolutional neural network. Secondly, we aim to apply our proposed methodology to the Hello Stretch robot to achieve real-time diagnosis of tomato leaf diseases. Thirdly, we assessed our method by comparing it to models like YOLOS, DETR, ViT, and Swin, demonstrating its ability to achieve state-of-the-art outcomes. For the purpose of the experiment, we used three datasets of tomato leaf diseases, namely KUTomaDATA, PlantDoc, and PlanVillage, where KUTomaDATA is being collected from a greenhouse in Abu Dhabi, UAE. Finally, we present a comprehensive analysis of the performance of our model and thoroughly discuss the limitations inherent in our approach. TomFormer performed well on the KUTomaDATA, PlantDoc, and PlantVillage datasets, with mean average accuracy (mAP) scores of 87\%, 81\%, and 83\%, respectively. The comparative results in terms of mAP demonstrate that our method exhibits robustness, accuracy, efficiency, and scalability. Furthermore, it can be readily adapted to new datasets. We are confident that our work holds the potential to significantly influence the tomato industry by effectively mitigating crop losses and enhancing crop yields.

\end{abstract}
\begin{figure*}[h]
\label{fig-block}
\centering
\includegraphics[width=\linewidth]{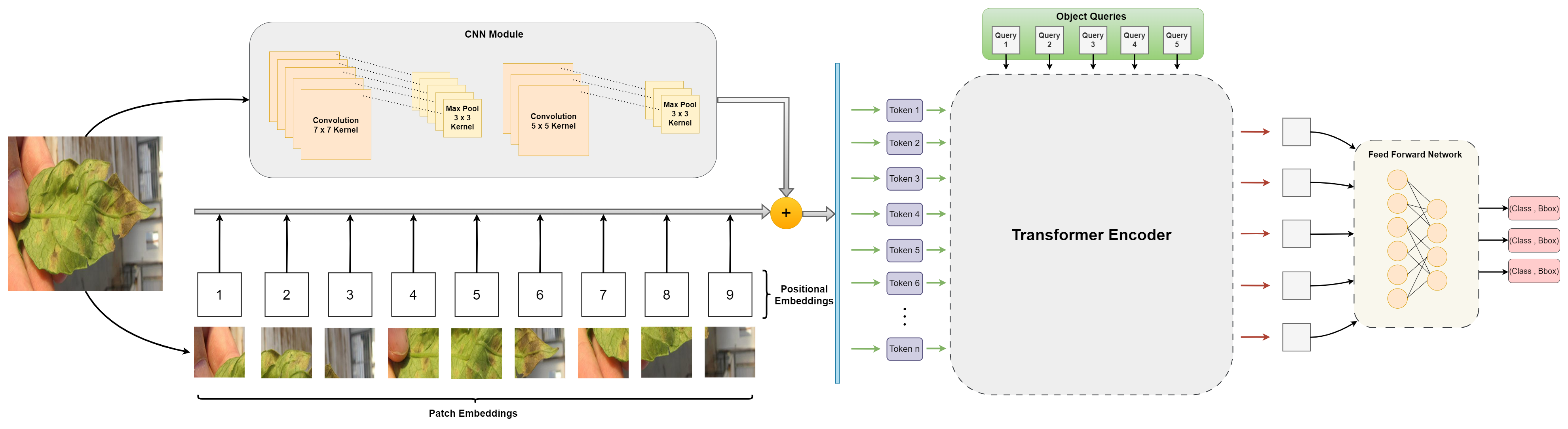}
\caption{A TomFormer architecture. It encompasses CNN and a Transformer encoder.}
\end{figure*}

\section{INTRODUCTION}

\label{sec:intro}
Solanum lycopersicum is the scientific name for tomatoes, which can grow on almost any well-drained soil \cite{pico1996viral} and is grown in fields by nine out of ten farmers. To use freshly produced tomatoes in their kitchens and enjoy excellent meals, many gardeners cultivate tomatoes in their home gardens. Depending on unfavourable seasonal and environmental conditions, plant diseases and pests significantly reduce plant yield, resulting in economic and societal losses. It takes time and money for people to identify pests and pathogens. Farmers still face difficulties in accurately identifying plant diseases. They are limited to speaking with other farmers or respective agricultural professionals as their only choices. The ability to recognize leaf diseases requires knowledge of plant diseases. As a result, farmers want automated AI image-based solutions.

The present state of computer vision applications, specifically in image and video analysis, involves utilising and acknowledging images as a dependable approach to disease diagnosis. This recognition is primarily facilitated through the accessibility of suitable software packages or tools. These technologies employ sophisticated image processing techniques, contributing to intelligent image identification, thereby enhancing recognition efficiency, cost reduction, and overall recognition accuracy~\cite{wang2020rin}.

Transformer networks have demonstrated their efficacy in various natural language processing tasks, such as machine translation, summarising texts, and question answering \cite{LIN2022111}. 
Recently, there has been a surge in the inclination towards employing transformer networks for computer vision tasks, encompassing image classification and object detection.
This research presents a novel methodology for identifying tomato leaf diseases using a customized transformer network integrated into the sophisticated Hello Stretch robot. Our methodology is founded upon the incorporation of the subsequent two fundamental concepts:

\begin{itemize}
    \item \textbf{Use a transformer network to extract features from images of tomato leaves.} 
    
    Transformer networks are ideally suited for this task because they can discover long-range dependencies in data sequences. This is important for tomato leaf disease detection, as the symptoms of different diseases often appear in different parts of the leaf.

    \item \textbf{Use the Hello Stretch robot to collect images of tomato leaves.} 
    
    The Hello Stretch robot is a mobile manipulator designed for indoor use. It has a depth camera to capture high-quality images of tomato leaves.
\end{itemize}


\section {Related Work}
\label{sec:related}
Computer vision has experienced rapid growth in recent years due to advancements in modern science and technology. This has led to a broader range of computer vision applications, including identifying and categorising plant diseases. 
Numerous artificial intelligence methods are currently employed for this purpose, encompassing a range of techniques such as k-nearest neighbors algorithm (K-NN), logistic regression (LR), decision trees (DTs), support vector machines (SVMs), and deep convolutional neural networks (DCNNs) \cite{Bharate2017ARO, jab-201803-0002}. 
These methods improve feature extraction when applied with image preprocessing. 
However, these methods are still weak regarding the model's efficiency.
Within the realm of supervised learning, CNNs can be regarded as comprehensive solutions for the purposes of classification or detection tasks. In their study, Brahimi et al. \cite{brahimi2017deep} employed a convolutional neural network (CNN) architecture to detect diseases in tomato leaves. 
Xibei et al.\cite{DBLP:journals/mta/HuangCZZWPYJ23} proposed in their work a network known as Fully Convolutional – Switchable Normalisation Dual Path Networks (FC-SNDPN) that was developed for autonomous recognition and detection of crop leaf diseases. Specifically, the focus was classifying eight types of diseases and insect pests commonly found on tomato leaves in southern China.
Using a spatial pyramid-oriented encoder–decoder cascade CNN architecture, Wang et al. \cite{hughes2015open} developed a method for detecting plant diseases in leaf tissue and segmenting the affected areas.

In recent years, CNNs have been effective in plant disease identification, as the studies discussed previously proved. Additionally, it has come to light that CNN-based architectures for detecting plant diseases predominantly take their cues from core frameworks that include Inception, GoogleNet, and ResNet\cite{YU2023100650}. In the domain of plant disease identification, a relatively new model known as the vision transformer (ViT) has recently been implemented \cite{borhani2022deep, alshammari2022olive}. 

The attention mechanism was incorporated into a deep residual CNN by Karthik et al. \cite{karthik2020attention} to detect infections in tomato leaves. 
Zeng et al. \cite{zeng2020crop} introduced a residual CNN structure enhanced with a self-attention mechanism to capture and extract relevant features from crop disease spots effectively. This approach was employed to identify and classify crop diseases accurately. 
In another study, for the purpose of disease identification in vegetables with complicated backgrounds, Zhou et al. \cite{zhou2021vegetable} developed a progressive learning network with an attention block.
Most recently, the research study conducted by Alshammari et al. \cite{alshammari2022olive} involved the development of a sophisticated hybrid model that integrates the visionary transformer architecture with the CNN architecture to ascertain the utmost efficacy and pertinence in the identification and categorization of olive diseases. 
The process of identifying plant diseases necessitates a high level of attentiveness toward the nuanced distinctions present within leaf images. Henceforth, we have amalgamated the Vision Transformer (ViT) with a CNN block to extract intricate features effectively. Additionally, we have introduced object queries to the encoder section of the transformer, following the implementation paradigm of the DETR architecture. The details are in the section \ref{sec:methods}.



\section{The  Proposed Method}
\label{sec:methods}
The \textbf{Tom}ato Trans\textbf{Former} (TomFormer) model, introduced as a novel object detection approach, draws inspiration from two influential models: Vision Transformer (ViT) \cite{dosovitskiy2020image} and DEtection TRansformer (DETR) \cite{carion2020end}. While DETR revolutionized object detection by replacing conventional two-stage methods with a single-stage transformer-based approach, ViT demonstrated the efficacy of transformers in image classification tasks. Building upon these advances, TomFormer incorporates critical modifications to address object detection requirements effectively and achieve superior performance in this domain. In TomFormer, the task of object detection is framed as a direct set prediction problem. Instead of generating bounding boxes through region proposal networks, TomFormer predicts the set of bounding boxes directly from the input image. It accomplishes this using a transformer encoder architecture with object queries as a side input. The encoder part of TomFormer processes the input image and extracts a set of feature maps. The positional embeddings allow the model to incorporate spatial information during the decoding process. To train TomFormer, it uses a bipartite matching loss and a Hungarian algorithm-based bipartite matching procedure to associate predicted boxes with ground-truth boxes. This loss ensures that each predicted box corresponds to a unique ground-truth box and vice versa, aiding in better learning and stable training.

\subsection{Image Processing Head}

In TomFormer, the cls token used in ViT for image classification is removed, and instead, N learnable object queries are introduced, enhancing the model's capability for object detection. During training, TomFormer adopts the bipartite matching loss from DETR, facilitating precise object detection. Moreover, TomFormer leverages both positional embeddings and convolutional features extracted by the CNN block, effectively merging low-level and high-level features for a comprehensive representation.

TomFormer exhibits a well-defined structure that effectively processes input images to facilitate object detection. The model's architecture consists of two pathways, wherein the input image undergoes distinct processing stages. The first pathway involves the utilization of a CNN block, which is responsible for extracting relevant features using convolutional layers. Subsequently, the output of the convolutional layer undergoes further refinement through max-pooling layers, where the dimension of each feature is reduced for enhanced computation. 

Following the successful extraction of critical features, the input is fused with patch embeddings of the input image. The original image is transformed into a series of 2D image patches to accomplish this, where the dimensions are represented as $x \in \mathbb{R}^{{P^2 \times C} \times N}$. where C is the number of channels, P is the dimension of a single image patch, and N is the number of resulting patches which is determined by $N = \frac{HW}{P^2}$, signifying the size of the input sequence of image patches for the subsequent transformer module.

Using a trainable linear projection, the reshaped patches are mapped to positional embeddings with dimensions D, making them fully compatible with the Transformer's constant latent vector size D across all layers
$E \in \mathbb{R}^{(P^2 \cdot C) \times D}$. The output of this projection is referred to as $x_{\text{PE}}$. These positional embeddings capture crucial positional information within the image patches, contributing to a comprehensive representation of the input data. Notably, the output of this projection is concatenated with the output from the CNN block, effectively merging the learned positional embeddings with the CNN-derived features. Remarkably, this strategic concatenation ensures a harmonious integration of both low-level and high-level visual information. This concatenated output is known with $x_{\text{PE + CNN}}$.

\subsection{TomFormer Encoder}
Following the successful merge of input features, they are further processed in the encoder block, where they are represented as a Token. This Token representation is a compact and informative encapsulation of the input image, facilitating subsequent object detection tasks. By incorporating these design principles, TomFormer adeptly combines the strengths of CNN-based feature extraction and the Transformer's attention mechanism, culminating in an efficient and robust model for object detection tasks.

In addition to the merging of positional embeddings and convolutional features, the TomFormer model introduces 20 randomly initialized learnable tokens, also referred to as object queries, represented as $x_{\text{oq}} \in \mathbb{R}^{20 \times D}$. These queries are appended to the combined input features, further enriching the representation for object detection. A practical rationale drives the choice of 20 tokens: in typical scenarios, the number of leaf objects in an image is expected to be relatively small up to 15 at max, and therefore, a small number of $x_{oq}$ is deemed sufficient to account for potential objects of interest. 

Specifically, the $x_{oq}$ tokens effectively serve as learnable representations of distinct tomato leaves that appear in the image. By introducing a set number of tokens, TomFormer intuitively balances model complexity and efficiency, ensuring that the architecture remains well-suited for object detection tasks with realistic object counts. The resulting sequence, denoted as $y_{\text{Res}}$, constitutes the enhanced input for the subsequent TomFormer encoder as shown in Equation~\ref{resultant}, enabling the model to effectively process and attend to the essential visual information necessary for accurate object detection. Through the thoughtful incorporation of these $x_{oq}$ tokens, TomFormer demonstrates its adaptability to handle various object detection scenarios with notable efficiency and robustness.

\begin{align}
y_{\text{Res}} &= \left[ x^1_{\text{PE + CNN}}; x^2_{\text{PE + CNN}}; x^3_{\text{PE + CNN}}; \ldots ; x^N_{\text{PE + CNN}} \right] \cup \nonumber \\
& \quad \left[ x^1_{\text{oq}}; x^2_{\text{oq}}; x^3_{\text{oq}}; \ldots ; x^{20}_{\text{oq}} \right]
\label{resultant}
\end{align}

In each encoder layer of the TomFormer, there are two fundamental components: the multi-head self-attention (MSA) block and the multi-layer perception (MLP) block. Both of these blocks are accompanied by LayerNorm (LN) \cite{ba2016layer} to normalize the intermediate results, and residual connections to facilitate information flow within the network~\cite{baevski2018adaptive , wang2019learning}. The MSA block allows the model to attend to different parts of the input sequence while capturing the long-range dependencies between elements. On the other hand, the MLP block consists of two hidden layers with the GELU~\cite{hendrycks2016gaussian} activation function, enabling the model to introduce non-linearity and enhance its capacity for complex pattern recognition in leaf images.

Formally, for the n-th TomFormer encoder layer, these components, namely MSA, LN, GELU-activated MLP, and residual connections, are combined to ensure efficient information propagation and effective feature extraction as shown in Equation~\ref{yn-1} and \ref{yn}, where n is the number of encoder layers. The use of these blocks within each encoder layer contributes to the overall expressive power of the TomFormer model and facilitates its ability to capture and learn intricate relationships and patterns in the images, ultimately leading to improved performance in various tasks and applications.
\begin{align}
y'_n &= \text{MSA}(\text{LN}(y_{n-1})) + y_{n-1}
\label{yn-1}
\end{align}
\begin{align}
y_n &= \text{MLP}(\text{LN}(y'_n)) + y'_n
\label{yn}
\end{align}

\subsection{Feed Forward Network}

The detector head of TomoFormer features a simplified and streamlined design reminiscent of the elegant image classification layer found in ViT. Classification and bounding box regression tasks are accomplished by utilizing a single feed-forward network (FFN) with separate parameters. This FFN consists of two hidden layers with intermediate ReLU activation functions. 
During fine-tuning of our datasets, a bipartite matching loss is introduced for each forward pass, establishing an optimal association between the predictions generated by object queries and the ground truth objects. This approach eliminates the need for reinterpreting ViT's output sequence into 2D feature maps for label assignment. Consequently, TomoFormer can perform object detection across any dimension without requiring precise spatial structure and geometry information as long as the input remains flattened to a consistent sequence format during each pass. This flexibility enhances the model's potential for diverse object detection tasks.


\subsection{The Hello Stretch robot \cite{stretch}}

\begin{figure}[h]
\centering
\includegraphics[scale=0.15]{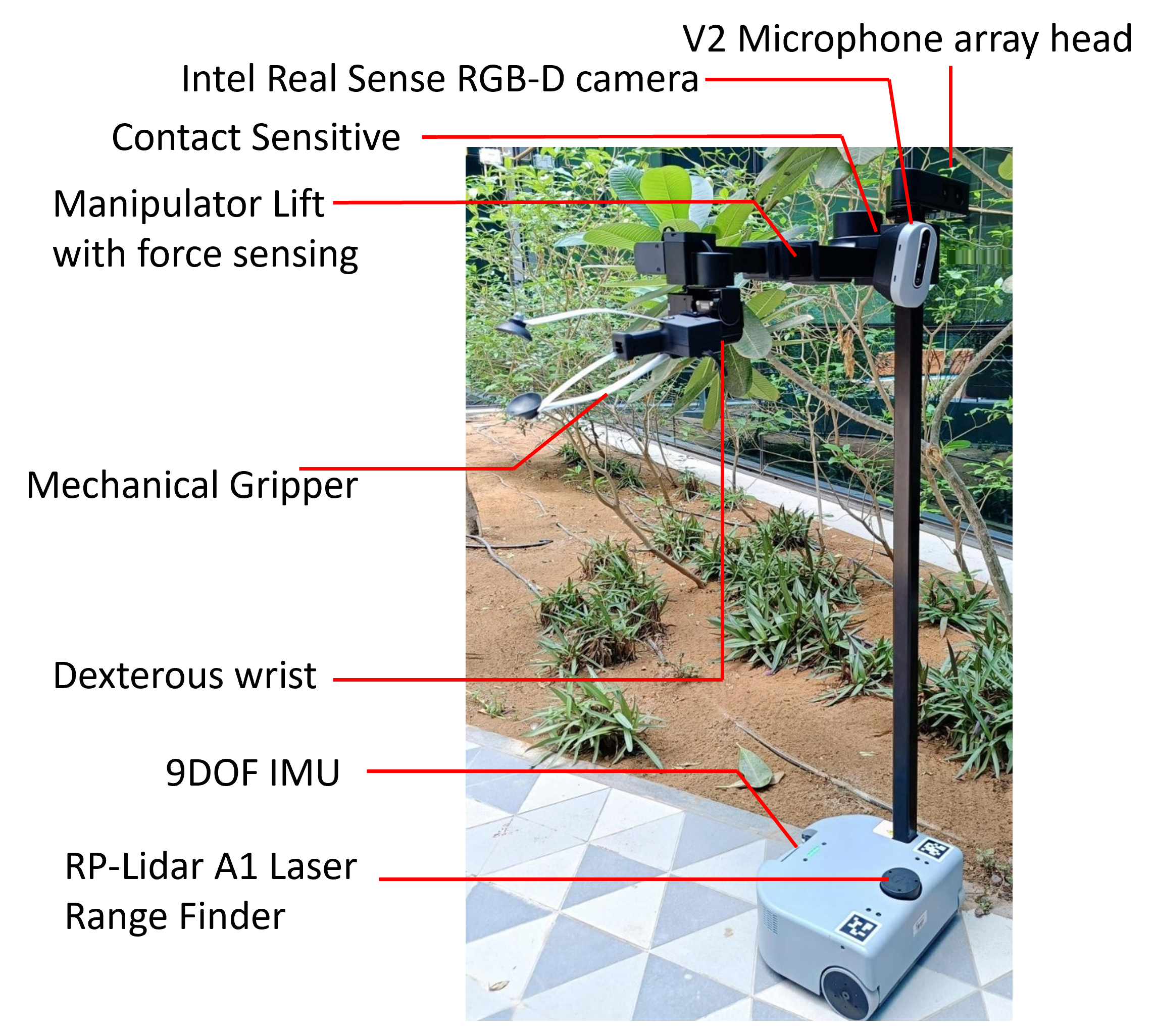}
\caption{A Hello stretch robot\cite{stretch}. It has mainly three parts, namely a Head, an Arm, and a Base. However, it has other components such as V2 microphone, Intel Real Sense RGB-D camera, Manipulator, IMU and RP-Lidar range finder. }
\label{fig-robot}
\end{figure}

The Hello stretch robot is a mobile manipulator designed for indoor use. It is a tall, thin robot with a single arm that extends from its head. The robot is made of a combination of aluminium and plastic, and it weighs 23 kg. The stretch hello robot is shown in Figure \ref{fig-robot}. 

The structure of the Hello Stretch robot is as follows:
\begin{itemize}
    \item Head: 
    The head of the robot houses the robot's computer, sensors, and actuators. The computer is a Jetson Nano, and the sensors include a depth camera, an inertial measurement unit (IMU), and a laser range finder. The actuators include the motors that drive the robot's wheels and the motor that drives the arm.

     \item Arm: 
    The arm of the robot is a telescoping arm that can reach up to 43 inches high and extend outward 20.5 inches. The arm is made of custom carbon fibre, and a single motor drives it. The arm has a gripper at the end that can hold objects up to 1.5 kg.

      \item Base: 
      The base of the robot is a two-wheeled mobile base. The base is made of aluminium, and it has a diameter of 34 cm. Two motors drive the base, which can move at a maximum speed of 0.6 m/s.
      
\end{itemize}
 
The Hello Stretch robot is versatile and can be used for various tasks, such as picking and placing objects, assembling products, and providing customer service. The robot is also open source, meaning it can be customized and further developed by the community.

The amalgamation of the Stretch Robot and the TomFormer model represents a significant advancement in agricultural technology, providing an intelligent, autonomous, and rapid disease monitoring system. With its ability to cover extensive areas and detect diseases early, the robot empowers farmers and agronomists with timely insights for targeted interventions and disease management strategies. The real-time feedback mechanism further enhances decision-making, as the robot promptly marks the affected areas on the plants, facilitating prompt and precise treatments. As we continue to explore the potential of this integrated system, its deployment in tomato detection showcases the transformative impact of robotics and artificial intelligence in revolutionizing the agricultural landscape.

\section{Experiments and Results}
\label{sec:results}
\begin{figure}[h]
\centering
\includegraphics[width=\linewidth]{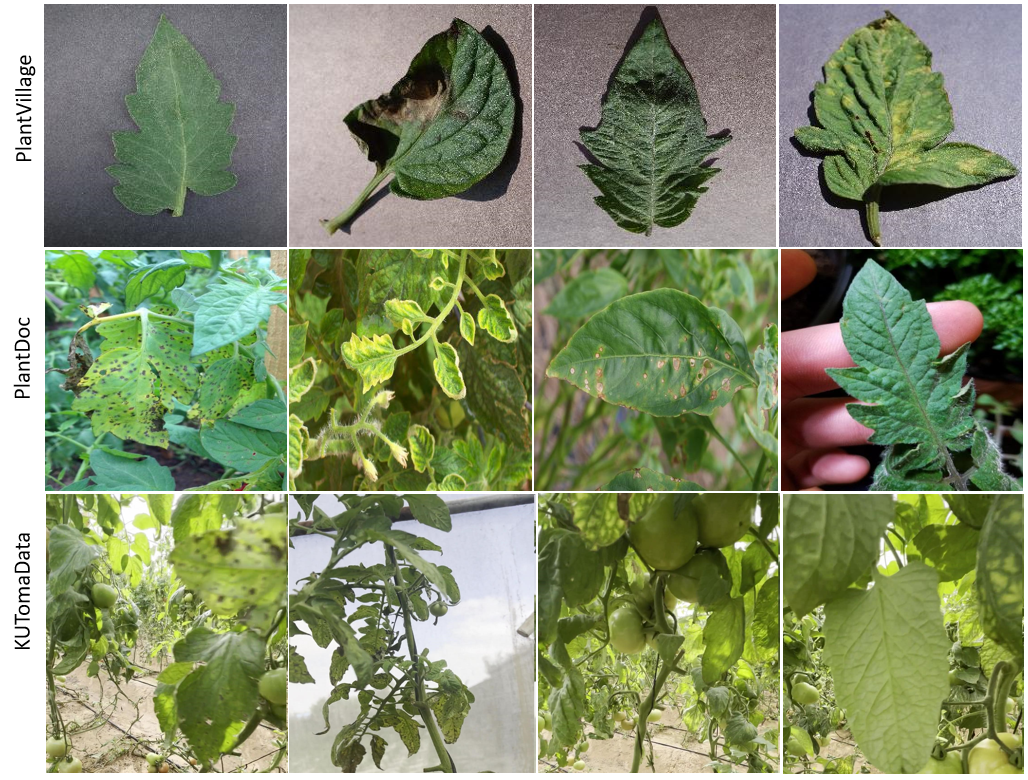}
\caption{The sample images from PlantVillage~\cite{DBLP:journals/corr/HughesS15}, PlantDoc~\cite{singh2020plantdoc} and KUTomaDATA datasets.}
\label{fig-data}
\end{figure}

We conduct a comprehensive analysis of the performance of our approach on multiple datasets, including our proprietary dataset and two publicly available datasets known as Plant Village~\cite{DBLP:journals/corr/HughesS15} and 
plantDoc~\cite{singh2020plantdoc} datasets. The sample images from each dataset are shown in Figure~\ref{fig-data}. 

We comprehensively analyse the results and perform a comparative evaluation of our method against state-of-the-art transformer-based approaches.
The performance evaluation of our proposed method in this study involved using the mean average precision ($mAP$) metric. 
The $mAP$ metric considers the balance between precision and recall while also considering the presence of false positives (FP) and false negatives (FN). This particular characteristic renders $mAP$ a suitable metric for various detection applications. The mathematical equation for $mAP$ is shown in Equation~\ref{miou}.

\begin{eqnarray}
mAP= \frac{1}{N}\sum_{i=1}^{N}AP_i
\end{eqnarray}\label{miou}


\subsection{Experimental Setup}

The proposed framework has been successfully implemented utilizing the PyTorch deep learning framework. The initial learning rate is set to 0.01 and is decreased by 5\% after each epoch. The experimental setup incorporated the following components: NVIDIA GeForce RTX 4090 Ti GPUs with a combined memory capacity of 24GB, the Ubuntu 20.04 operating system, an Intel i9 CPU, and 64GB of RAM. The proposed model comprises 1.30 million network parameters, with an estimated inference time of approximately 200 seconds. After training our model on the aforementioned system, we implemented it on the stretch robot for real-time inference. The stretch robot used Robot Operating System (ROS) and python3. \AK{The seamless integration of the TomFormer model into the software environment of the Hello Stretch Robot was instrumental for effective tomato disease detection. We ensured that all the necessary dependencies for the TomFormer model were installed. Subsequently, we created a dedicated interface within the ROS environment. This interface allowed us to receive image data from the robot's built-in depth camera. These images were then passed to the TomFormer model for inference. The process continued with the model being loaded using Python3. Once loaded, the model executed the inference process, which resulted in accurately detecting tomato diseases within the images. The output images containing detected objects are then saved in the robot repository. This integration and inference workflow was pivotal in enhancing the robot's capabilities for real-time disease management in tomato fields. }
\begin{figure}[h]
\centering
\includegraphics[scale=0.3]{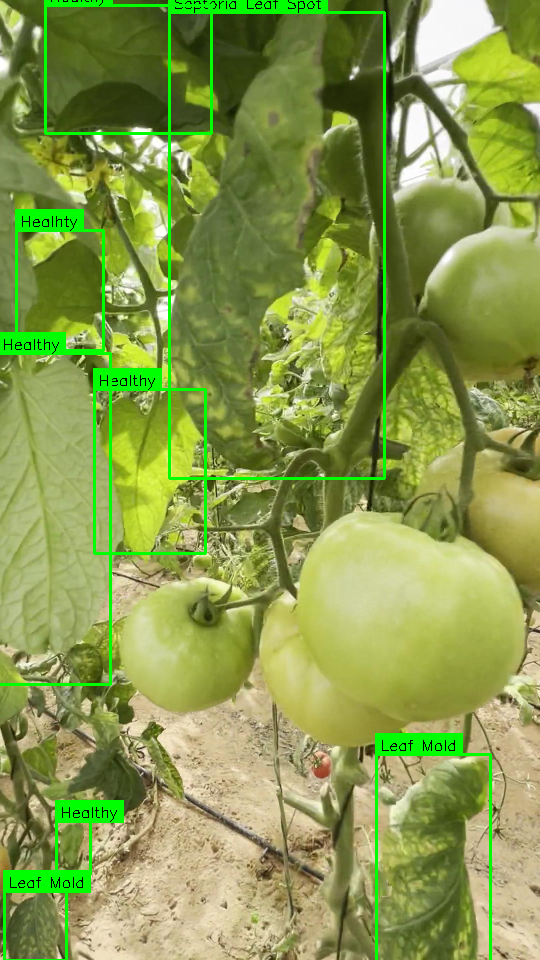}
\caption{Tomato leaf disease detection using Hello stretch robot camera}
\label{output}
\end{figure}

Figure~\ref{output} shows the inference results displayed on the monitor for an image captured by a stretch robot.

\subsection{KUTomaDATA Dataset} 
The dataset used in this study comprises 939 images of tomatoes captured within greenhouses in Al Ajban, Abu Dhabi, United Arab Emirates. These images were acquired using mobile phone cameras and encompass a wide range of leaf images, ranging from healthy ones to those affected by various diseases. To ensure diversity and representation of different disease categories, the dataset has been meticulously partitioned into eight distinct classes, which were determined based on the visual appearance of each disease from publicly available datasets, i.e., PlantDoc~\cite{singh2020plantdoc}, PlantVillage~\cite{DBLP:journals/corr/HughesS15}.
The specific classes included in the KUTomaDATA dataset are as follows: healthy, bacterial spots, early blight, late blight, leaf mold, septoria leaf spot, mosaic virus, and yellow leaf curl. \AK{The dataset consists of varying numbers of images for each class, with 118, 113, 122, 109, 116, 119, 124, and 118 images, respectively.}
\AK{As a result, each image underwent labelling using the Roboflow~\cite{Roboflow} annotator, and the resulting annotations were exported in JSON format. These annotations include the object labels, coordinates, and image dimensions.}


\subsection{PlantDoc Dataset~\cite{singh2020plantdoc}}

The dataset was created through the annotation of publicly available images, requiring a total of 300 human hours. It consists of a total of 2,598 data points, encompassing 13 different plant species and up to 17 types of diseases. For our study, we picked 700 leaf images of healthy and disease classes for the tomato dataset to be used for training and testing. For scientists and programmers engaged in plant disease identification, the PlantDoc dataset is a valuable resource. The dataset is vast and varied, and it includes information on many different plant species and ailments. Additionally, the dataset is well-annotated, making it simple to use for developing and testing a variety of models.

\subsection{PlantVillage Dataset~\cite{DBLP:journals/corr/HughesS15}}

The tomato disease images were sourced from the Plant Village dataset~\cite{DBLP:journals/corr/HughesS15}, which comprises a collection of over 50,000 images representing 14 distinct crops, such as tomatoes, potatoes, grapes, apples, corn, blueberries, raspberries, soybeans, squash, and strawberries. These images were captured under carefully controlled conditions. From this dataset, we separated eight types of tomato leaf diseases: bacterial spots, early blight, late blight, leaf mold, Septoria leaf spot, mosaic virus, yellow leaf curl virus, and healthy. The experimental dataset consists of 700 images in total. 

\begin{table*}[htbp]
\centering
\caption{A comparative analysis of different object detection models based on each class's mean Average Precision (mAP).}
\label{tab:method_comparison}
\adjustbox{max width=\textwidth}{

\begin{tabular}{cccccccccccccccc}
\toprule
\textbf{\huge Class} & \multicolumn{3}{c}{\textbf{\huge TomFormer}} & \multicolumn{3}{c}{\textbf{\huge YOLOS \cite{fang2021you}}} & \multicolumn{3}{c}{\textbf{\huge DETR} \huge \cite{carion2020end}} & \multicolumn{3}{c}{\textbf{\huge ViT} \huge \cite{dosovitskiy2020image}} & \multicolumn{3}{c}{\textbf{\huge Swin} \huge \cite{liu2021swin}} \\
\cmidrule(lr){2-4} \cmidrule(lr){5-7} \cmidrule(lr){8-10} \cmidrule(lr){11-13} \cmidrule(lr){14-16}
& \huge KUTomaDATA & \huge PlantDOC & \huge PlantVillage  & \huge KUTomaDATA & \huge PlantDOC & \huge PlantVillage & \huge KUTomaDATA & \huge PlantDOC & \huge PlantVillage & \huge KUTomaDATA & \huge PlantDOC & \huge PlantVillage & \huge KUTomaDATA & \huge PlantDOC & \huge PlantVillage \\
\hline
\midrule
\textbf{\huge Healthy leaf} & $\textbf{\huge 92\%}$ & $\textbf{\huge 88\%}$ & $\textbf{\huge 90\%}$ & \huge $84\%$ & \huge $82\%$ & \huge $85\%$ & \huge $87\%$ & \huge $84\%$ & \huge $88\%$ & \huge $79\%$ & \huge $78\%$ & \huge $82\%$ & \huge $85\%$ & \huge $83\%$ & \huge $87\%$ \\

\textbf{\huge Bacterial spots} & $\textbf{\huge 90\%}$ & $\textbf{\huge 81\%}$ & $\textbf{\huge 84\%}$ & \huge $81\%$ & \huge $77\%$ & \huge $80\%$ &\huge  $83\%$ & \huge $80\%$ & \huge $82\%$ & \huge $75\%$ & \huge $69\%$ & \huge $73\%$ & \huge $79\%$ & \huge $72\%$ & \huge $76\%$ \\

\textbf{\huge Early blight} & $\textbf{\huge 80\%}$ & $\textbf{\huge 76\%}$ & $\textbf{\huge 78\%}$ & \huge $74\%$ & \huge $73\%$ & \huge $76\%$ & \huge $79\%$ & \huge $76\%$ & \huge $77\%$ & \huge $67\%$ & \huge $66\%$ & \huge $69\%$ & \huge $72\%$ & \huge $70\%$ & \huge $74\%$ \\

\textbf{\huge Late blight} & $\textbf{\huge 88\%}$ & $\textbf{\huge 84\%}$ & $\textbf{\huge 86\%}$ & \huge $81\%$ & \huge $80\%$ & \huge $82\%$ & \huge $84\%$ & \huge $82\%$ & \huge $84\%$ & \huge $75\%$ & \huge $74\%$ & \huge $77\%$ & \huge $79\%$ & \huge $78\%$ & \huge $82\%$ \\

\textbf{\huge Leaf mold} & $\textbf{\huge 86\%}$ & $\textbf{\huge 82\%}$ & $\textbf{\huge 84\%}$ &\huge  $80\%$ & \huge $78\%$ & \huge $81\%$ & \huge $82\%$ & \huge $80\%$ & \huge $83\%$ & \huge $72\%$ &\huge  $71\%$ &\huge  $75\%$ &\huge  $76\%$ & \huge $75\%$ &\huge  $79\%$ \\

\textbf{\huge Septoria leaf spot} & \huge $\textbf{90\%}$ & \huge $\textbf{78\%}$ & \huge $\textbf{80\%}$ & \huge $83\%$ & \huge $75\%$ & \huge $78\%$ & \huge $86\%$ & \huge $76\%$ & \huge $80\%$ & \huge $76\%$ & \huge $69\%$ & \huge $72\%$ & \huge $81\%$ & \huge $72\%$ & \huge $76\%$ \\

\textbf{\huge Mosaic virus} & $\textbf{\huge 87\%}$ & $\textbf{\huge 83\%}$ & $\textbf{\huge 85\%}$ & \huge $82\%$ & \huge $80\%$ & \huge $83\%$ & \huge $84\%$ & \huge $82\%$ & \huge $86\%$ & \huge $74\%$ & \huge $73\%$ & \huge $76\%$ & \huge $78\%$ &\huge  $77\%$ & \huge $81\%$ \\

\textbf{\huge Yellow leaf curl} & $\textbf{\huge 84\%}$ & $\textbf{\huge 80\%}$ & $\textbf{\huge 82\%}$ & \huge $78\%$ & \huge $77\%$ & \huge $79\%$ & \huge $81\%$ & \huge $79\%$ & \huge $82\%$ & \huge $71\%$ & \huge $70\%$ & \huge $74\%$ & \huge $75\%$ & \huge $73\%$ & \huge $77\%$ \\

\hline
\textbf{\huge Average} & $\textbf{\huge 87\%}$ & $\textbf{\huge 81\%}$ & $\textbf{\huge 83\%}$ & \huge $80\%$ &\huge  $77\%$ & \huge $80\%$ & \huge $82\%$ & \huge $79\%$ & \huge $83\%$ & \huge $73\%$ & \huge $71\%$ & \huge $75\%$ &\huge  $78\%$ & \huge $76\%$ & \huge $80\%$ \\

\hline
\end{tabular}
}
\end{table*}
\section{Discussion}
The comprehensive evaluation of the object detection models across the three class labels reveals important insights for tomato leaf disease detection. TomFormer consistently emerged as the top-performing model across all three datasets (KUTomaDATA, PlantDoc~\cite{singh2020plantdoc}, PlantVillage~\cite{DBLP:journals/corr/HughesS15}), indicating its robustness and efficacy in detecting and localizing plant diseases and healthy leaf conditions. Detailed results are presented in the later parts of this section. However, it is essential to consider various factors, such as model complexity, computational resources, and real-world applicability, when selecting the most suitable model for a specific plant disease identification scenario. 

\AK{The computational capabilities of the Hello Stretch Robot are of paramount significance within our integrated framework. The robot is equipped with a high-performance computing unit with a multi-core Intel i5-8259U processor paired with 16GB of RAM. This configuration offers a substantial computational capacity, rendering it well-suited for tasks including image processing and the execution of inference for object detection.
}

The competitive performance of YOLOS \cite{fang2021you}, DETR \cite{carion2020end}, ViT \cite{dosovitskiy2020image}, and Swin transformer \cite{liu2021swin} underscores their potential as viable alternatives for tomato disease detection. Their comparative results are presented in the Table~\ref{tab:method_comparison}.
\label{sec:discuss}

\subsection{Results on KUTomaDATA}
In evaluating object detection models on the KUTomaDATA classes, which comprise a diverse set of plant health conditions and diseases, we observed varying performance levels across mAP. TomFormer emerged as the top-performing model with an impressive mAP score of 87\%. 
The demonstrated outcome showcases TomFormer's remarkable ability to effectively identify and pinpoint various plant diseases, such as bacterial spots, early blight, late blight, leaf mold, septoria leaf spot, mosaic virus, yellow leaf curl, and healthy leaves within the KUTomaDATA. The mAP score of 80\% achieved by YOLOS \cite{fang2021you}, 82\% by DETR \cite{carion2020end}, 73\% by ViT \cite{dosovitskiy2020image}, and 77\% by Swin \cite{liu2021swin} demonstrates their competence in detecting and classifying plant health conditions. However, TomFormer's superior mAP score reaffirms its effectiveness in handling the complexities and diversities inherent in this class.

\subsection{Results on PlantDOC}
The evaluation results provided further insights into the models' performances in the context of the PlantDoc dataset, which encompasses a subset of plant healthy and disease classes. TomFormer demonstrated promising performance with an mAP score of 81\%, indicating its ability to accurately detect and classify bacterial spots, early blight, late blight, leaf mold, septoria leaf spot, and healthy leaves. Similarly, the models YOLOS \cite{fang2021you}, DETR \cite{carion2020end}, ViT \cite{dosovitskiy2020image}, and Swin \cite{liu2021swin} achieved competitive mAP scores of 77\%, 79\%, 71\%, and 76\%, respectively. The close proximity of the mAP scores suggests that all models possess the competence to address the complexities present in the PlantDoc class effectively. Nevertheless, TomFormer's edge in performance reiterates its potential as a strong contender for tomato disease detection on this dataset. 

\subsection{Results on PlantVillage}
The PlantVillage class, characterized by a broad spectrum of tomato leaf images with uniform and simple backgrounds, was the subject of our final evaluation. TomFormer, once again, exhibited exceptional performance, achieving the highest mAP score of 83\%. This outcome highlights TomFormer's proficiency in accurately detecting and localizing various tomato diseases in a uniform environment. While the other models, YOLOS \cite{fang2021you}, DETR \cite{carion2020end}, ViT \cite{dosovitskiy2020image}, and Swin \cite{liu2021swin}, achieved competitive mAP scores of 80\%, 83\%, 75\%, and 80\%, respectively, TomFormer's consistent superiority in this class underscores its versatility and adaptability in handling the images with uniform background.


\subsection{Critical Evaluation}
The mAP scores for each class vary significantly across different models. For instance, the performance of TomFormer and YOLOS \cite{fang2021you} is consistently higher than DETR \cite{carion2020end}, ViT \cite{dosovitskiy2020image}, and Swin \cite{liu2021swin} for most of the classes. This indicates that these models are more suitable for classes except for early blight and septoria leaf spots as these classes consistently show lower mAP scores across all models. This suggests that these classes might be more challenging to detect and classify accurately, which is mostly due to their visual similarity with other classes. Upon analyzing the table, a notable trend emerges, indicating that all the models, except for TomFormer, demonstrated excellent performance on the PlantVillage dataset as compared to the PlantDoc and KUTomaDATA. This observation aligns with the dataset's characteristics, as it consists of leaf images with a uniform background. The uniformity in background images facilitates easier object detection and classification, leading to higher mAP values for the models. The higher mAP scores achieved by the other models on the PlantVillage dataset validate their effectiveness in the accurate detection of plant diseases in scenarios where the visual appearance of leaves is consistent and well-defined. However, TomFormer, despite its exceptional performance on KUTomaDATA, exhibits a relatively lower mAP score on the PlantVillage dataset. This result may be attributed to the dataset's specific challenges, where TomFormer faces difficulty distinguishing between diseases with subtle visual differences due to the uniform background setting. Overall, this comparison underscores the importance of using domain-specific datasets like PlantVillage, which mimic real-world scenarios and provide a robust evaluation of object detection models' performance in practical disease diagnosis.

\begin{figure}[h]
\label{fig-block}
\centering
\includegraphics[width=\linewidth]{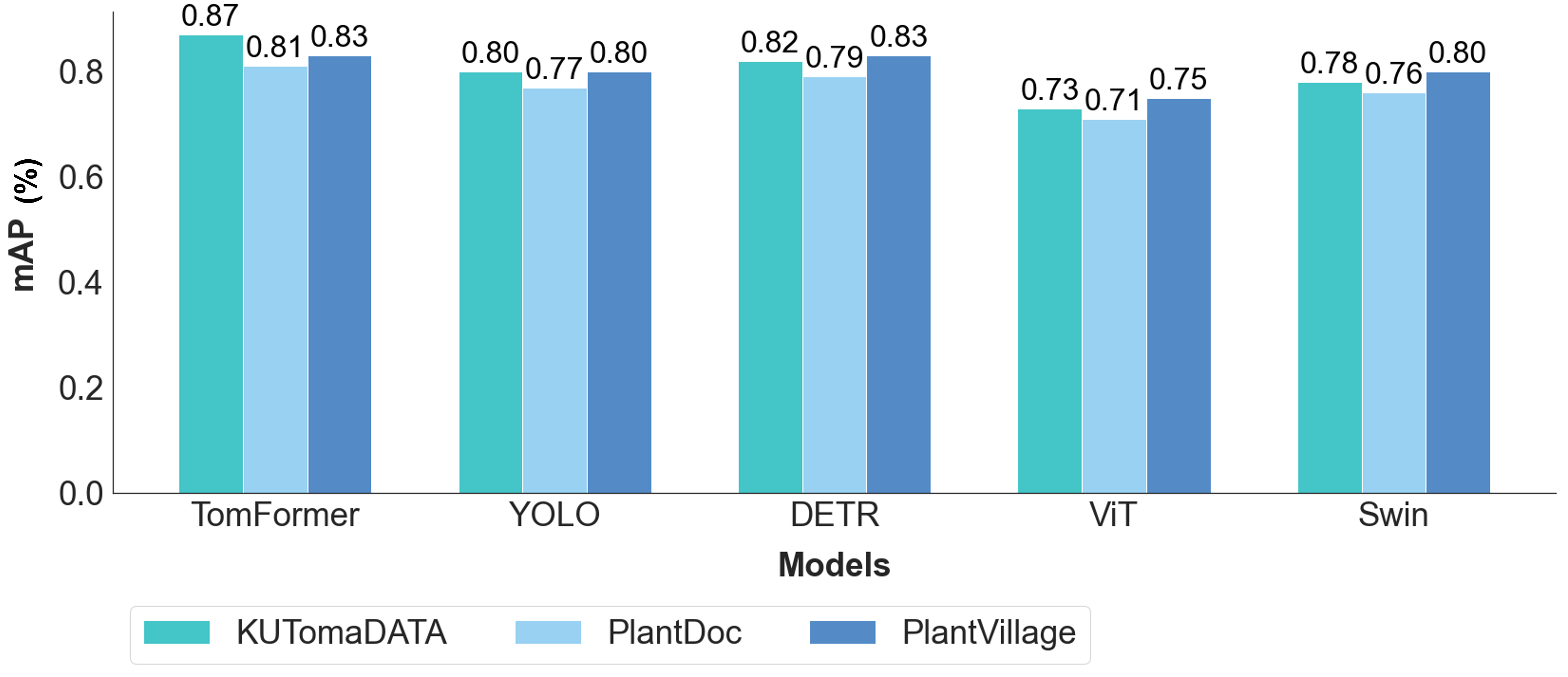}
\caption{Visualization of mean average precision for each model on TPlantVillage~\cite{DBLP:journals/corr/HughesS15}, PlantDoc~\cite{singh2020plantdoc} and KUTomaDATA datasets.}
\end{figure}

\section{Conclusion}
\label{sec:coclusion}
In conclusion, this paper introduces TomFormer, a transformer-based model designed for detecting diseases in tomato leaves. It combines a visual transformer and a convolutional neural network to provide an innovative approach to disease detection. The Hello Stretch robot, which uses TomFormer, can diagnose tomato leaf diseases in real time, making it a practical agricultural solution. Additionally, the inclusion of the KUTomaDATA dataset expands the research area. Extensive experiments and comparisons with other transformer models demonstrate TomFormer's robustness, accuracy, efficiency, and scalability, with mAP scores of 87\%, 81\%, and 83\% on the KUTomaDATA, PlantDoc, and PlantVillage datasets, respectively. This work has the potential to significantly benefit the tomato industry by reducing crop losses and improving yields, offering an effective tool for early disease detection and promoting sustainable agricultural practices.












\end{document}